\documentclass[apj]{emulateapj}

\newcommand{\kms}{km~s$^{-1}$}
\newcommand{\mgii}{\textrm{Mg}~\textsc{ii}}
\newcommand{\nev}{[\textrm{Ne}~\textsc{v}]}
\newcommand{\oii}{[\textrm{O}~\textsc{ii}]}
\newcommand{\hii}{\textrm{H}~\textsc{ii}}
\newcommand{\oiii}{[\textrm{O}~\textsc{iii}]}
\newcommand{\msun}{M$_{\odot}$}
\newcommand{\mstar}{M$_{*}$}
\newcommand{\units}{M$_{\odot}$~yr$^{-1}$~kpc$^{-2}$}
\newcommand{\lrest}{\lambda_{\textnormal{\scriptsize{rest}}}}
\newcommand{\lobs}{\lambda_{\textnormal{\scriptsize{obs}}}}
\newcommand{\sigmasfr}{\Sigma_{\textnormal{\scriptsize{SFR}}}}
\newcommand{\sfrir}{\textnormal{SFR}_{\textnormal{\scriptsize{IR}}}}

\shorttitle{Compact Starbursts Driving High-Velocity Outflows}
\shortauthors{Diamond-Stanic et al.}

\begin{document}
\slugcomment{Accepted for publication in ApJ Letters, 2012 July 18}

\title{High-Velocity Outflows Without AGN Feedback:
  \\ Eddington-Limited Star Formation in Compact Massive Galaxies}

\author{Aleksandar M. Diamond-Stanic\altaffilmark{1,2}, John
  Moustakas\altaffilmark{1}, Christy A. Tremonti\altaffilmark{3},
  Alison L. Coil\altaffilmark{1}, Ryan C. Hickox\altaffilmark{4}, Aday
  R. Robaina\altaffilmark{5}, Gregory H. Rudnick\altaffilmark{6}, \&
  Paul H. Sell\altaffilmark{3} }

\altaffiltext{1}{Center for Astrophysics and Space Sciences,
  University of California, San Diego, La Jolla, CA 92093, USA}
\altaffiltext{2}{Center for Galaxy Evolution Fellow; aleks@ucsd.edu}
\altaffiltext{3}{Department of Astronomy, University of
  Wisconsin-Madison, Madison, WI 53706, USA}
\altaffiltext{4}{Department of Physics and Astronomy, Dartmouth
  College, Hanover, NH 03755, USA} 

\altaffiltext{5}{Institut de Ci{\'e}ncies del Cosmos, University of
  Barcelona, 08028 Barcelona, Spain}

\altaffiltext{6}{Department of Physics and Astronomy, University of
  Kansas, Lawrence, KS 66045, USA}

\begin{abstract}

We present the discovery of compact, obscured star formation in
galaxies at $z\sim0.6$ that exhibit $\gtrsim1000$~\kms\ outflows.
Using optical morphologies from the Hubble Space Telescope and
infrared photometry from the Wide-field Infrared Survey Explorer, we
estimate star formation rate (SFR) surface densities that approach
$\sigmasfr\approx3000$~\units, comparable to the Eddington limit from
radiation pressure on dust grains.  We argue that feedback associated
with a compact starburst in the form of radiation pressure from
massive stars and ram pressure from supernovae and stellar winds is
sufficient to produce the high-velocity outflows we observe, without
the need to invoke feedback from an active galactic nucleus.

\end{abstract}


\section{Introduction}

The central regions of elliptical galaxies are thought to form in
compact starbursts \citep{kor09,hop09}.  Feedback associated with such
starbursts can produce outflows driven by thermal energy from
supernova explosions \citep{che85}, stellar winds \citep{lei92}, and
momentum input from both supernova ram pressure and radiation pressure
on dust grains \citep{mur05}.  It has been argued that such feedback
imposes a limit on the maximum star-formation rate (SFR) surface
density ($\sigmasfr$) for starbursts \citep{leh96,meu97,mur05,tho05}
and the maximum stellar surface density for elliptical galaxies and
star clusters \citep{hop10}.

Galactic winds are ubiquitous in star-forming galaxies at all
redshifts and generally exhibit outflow velocities in the
100--500~\kms\ range, which can be attributed to the stellar processes
described above \citep{hec00,sha03,mar05,rup05,wei09,rub10}.  Outflows
with significantly higher velocities ($|v|>1000$~\kms) were discovered
by \citet{tre07} in a sample of massive
($\textnormal{\mstar}\approx10^{11}$~\msun) post-starburst galaxies at
$z\sim0.6$, and it was suggested that a more energetic source such as
feedback from an accreting supermassive black hole \citep{sil98,dim05}
may be responsible for launching the winds \citep[see][for a recent
  review]{fab12}.

However, it is also plausible that feedback from a compact starburst
could expel gas with such large velocities.  Indeed, there is evidence
for a positive correlation between outflow velocity and starburst
luminosity \citep{mar05,rup05,tre07}, albeit with significant scatter.
Furthermore, \citet{hec11} recently found outflows with maximum
velocities reaching $1500$~\kms\ in a sample of local starbursts with
compact nuclei, and argued that such velocities could be explained by
a wind launched from $r_0\sim100$~pc and driven by feedback from
massive stars and supernovae.

In this Letter, we measure sizes and SFRs for a sample of massive
galaxies at $z\sim0.6$ that exhibit $\gtrsim1000$~\kms\ outflows,
expanding on the initial study by \citet{tre07}.  We seek to test
whether the energetic outflows in these galaxies could have been
driven by feedback from starbursts with very large SFR surface
densities.  Our analysis combines galaxy sizes measured with the
Hubble Space Telescope (HST) with SFRs and stellar masses estimated
from Wide-field Infrared Survey Explorer (WISE), Spitzer Space
Telescope, Sloan Digital Sky Survey (SDSS), and Galaxy Evolution
Explorer (GALEX) photometry.

\begin{figure}[!t]
\begin{center}
\includegraphics[angle=0,scale=0.41]{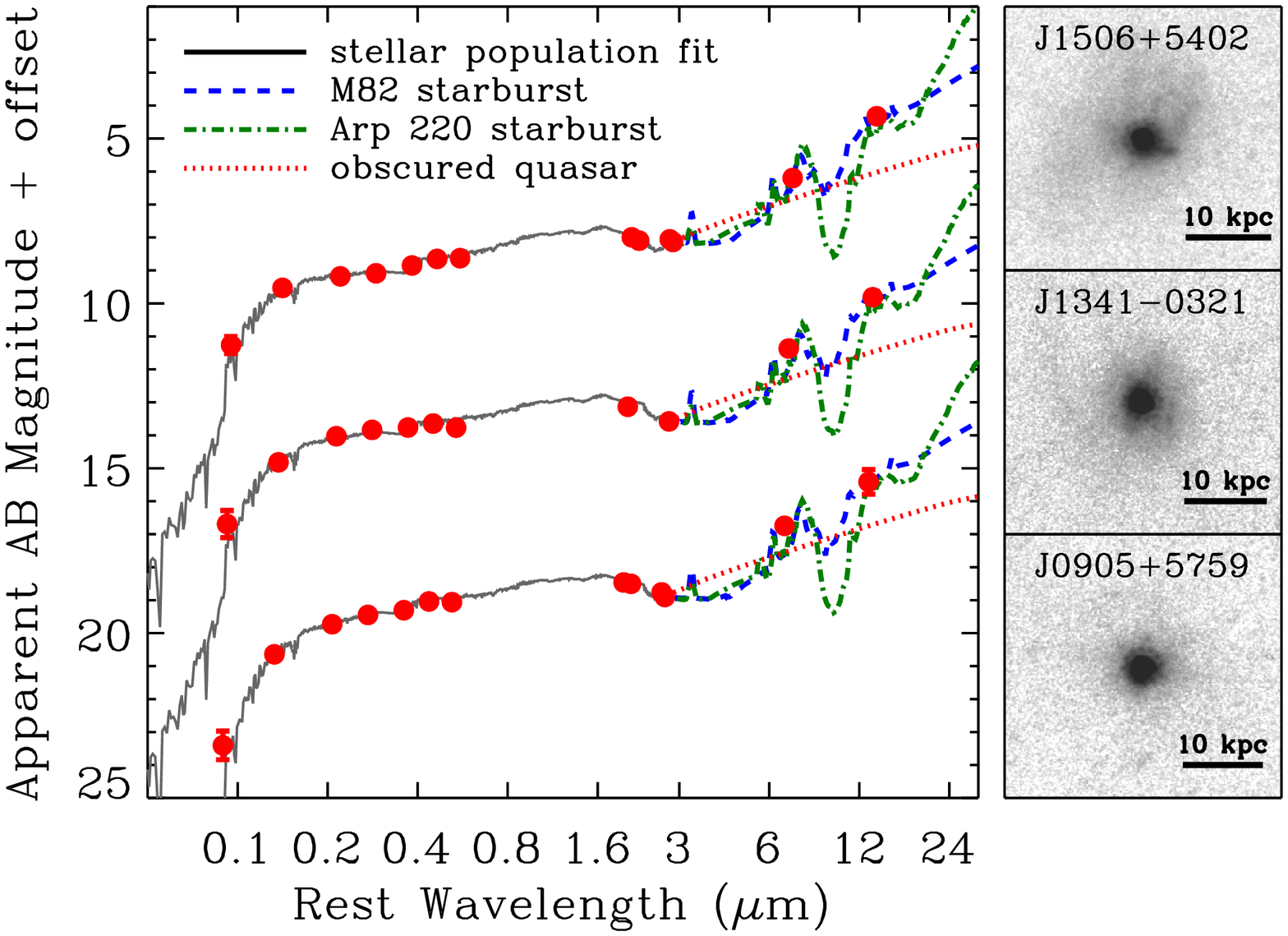}
\caption{Left: Observed UV--IR SEDs ($\lrest=0.1$--15~$\mu$m) for
  three galaxies with the largest SFR surface densities
  ($\sigmasfr\approx3000$~\units).  The top two SEDs are offset by 5
  and 10 magnitudes, respectively.  We show stellar population fits to
  the $\lrest=0.1$--3~$\mu$m emission (black solid line) and three
  templates for dust emission \citep[M82 starburst, blue dashed line;
    Arp 220 starburst, green dot-dashed line; obscured quasar, red
    dotted line;][]{pol07} scaled to match the $\lobs=4.6~\mu$m band.
  The starburst templates provide reasonable fits to the $\lobs=12$
  and 22~$\mu$m WISE photometry, while the quasar template does not.
  Right: HST/WFC3 F814W images (probing $\lrest\approx5000$~\AA)
  showing that these galaxies are dominated by a compact nucleus.}
\label{fig:seds}
\end{center}
\end{figure}

\section{Analysis}

\subsection{Morphologies and Sizes}

We observed 29 galaxies with HST (programs 12019 and 12272) selected
from a parent sample of $\sim10^3$ galaxies at $0.35<z<1.0$ with
post-starburst spectral features: B or A-star dominated stellar
continua and moderately weak nebular emission
($\textnormal{EW}(\oii)<20$~\AA; see C. Tremonti, et al., in
preparation for more details).  The galaxies targeted with HST were
those with the youngest derived post-burst ages,
$t_{burst}\lesssim300$~Myr.  Therefore this sample has bluer $U-B$
colors and stronger emission lines than typical post-starburst samples
\citep{coi11}.  Our subsequent UV--IR SED analysis (see
Section~\ref{sec:seds}) reveals significant ongoing star formation
($\textnormal{SFR}>50$~\msun~yr$^{-1}$) in the 14/29 galaxies with
WISE 22~$\mu$m detections, calling into question the post-starburst
nature of roughly half of the HST sample.

Using the F814W filter on the WFC3/UVIS channel, which has
$0.04\arcsec$ pixels, we obtained $4\times10$~min exposures in a
single orbit for each galaxy.  The dithered images were processed with
MultiDrizzle\footnote{http://stsdas.stsci.edu/multidrizzle/} to
produce science mosaics with $0.02\arcsec$ pixels.  For each galaxy,
we use GALFIT \citep{pen02} to model the two-dimensional surface
brightness profile with a single Sersic component (characterized by
Sersic index $n$ and effective radius $r_e$), using stars in the
images to construct the model point-spread function (PSF).  In cases
where the best-fit model returns $n>4$, we also fit an $n=4$ de
Vaucouleurs model, yielding a larger $r_e$ value (due to the
covariance between $n$ and $r_e$); we use these larger effective
radii in our analysis.

In this Letter, we highlight the galaxies with the smallest $r_e$ and
largest $\sigmasfr$ values because such extreme starbursts could
conceivably produce the high-velocity outflows we observe (see
Section~\ref{sec:discussion}).  We show HST images for the three
highest $\sigmasfr$ galaxies in Figure~\ref{fig:seds}.  In all three
cases, the single-component GALFIT model accounts for $>85\%$ of the
total flux.  The residuals show diffuse emission that is consistent
with these systems being late-stage galaxy mergers, although we defer
a detailed study of the merger stage to future work.

\begin{figure}[!t]
\begin{center}
\includegraphics[angle=0,scale=0.41]{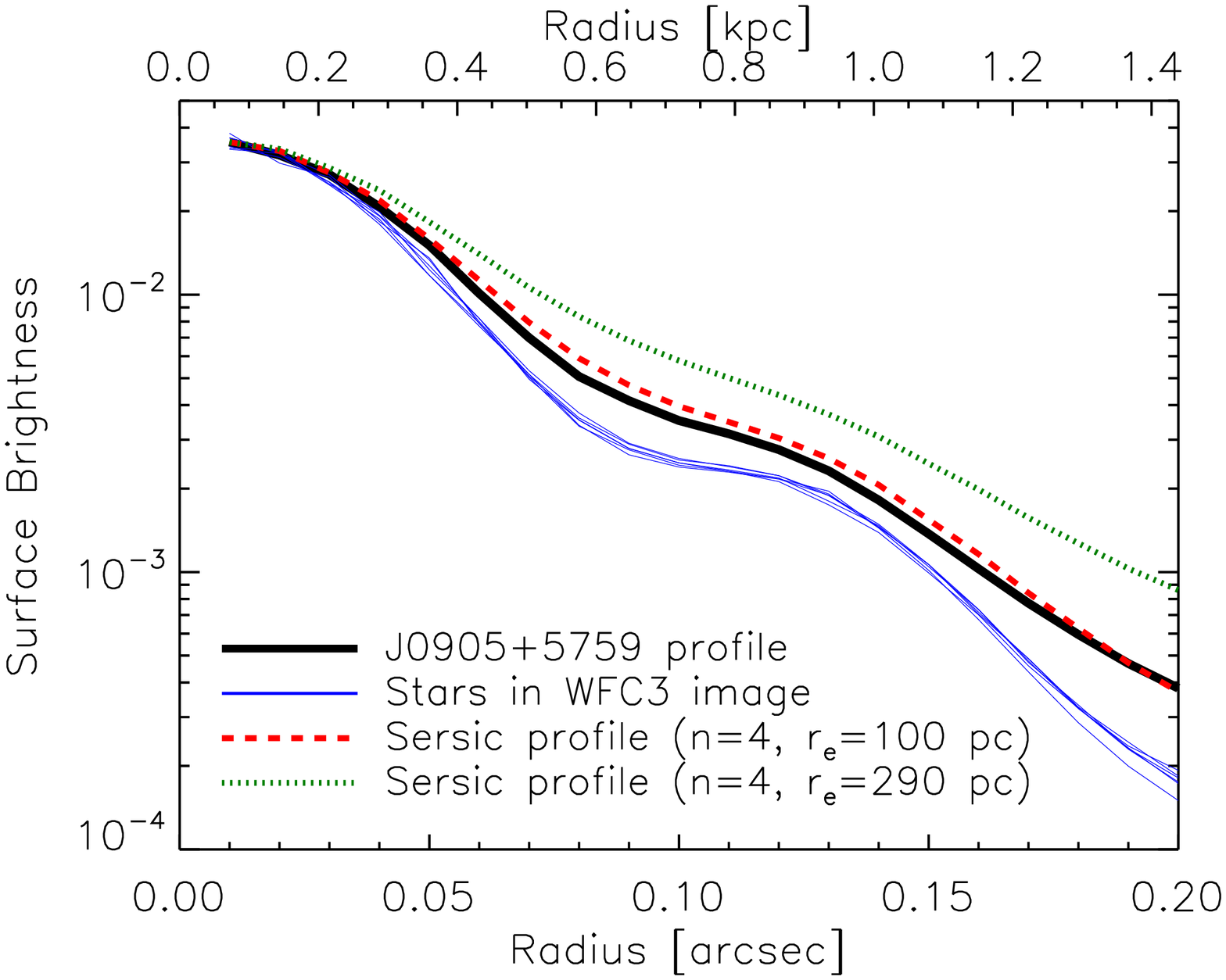}
\caption{One-dimensional surface brightness profile for J0905+5759,
  which has the smallest effective radius in the sample.  The observed
  profile is shown as the solid black line, and the profiles of six
  stars in the same image are shown in blue, normalized to the same
  central surface brightness.  The best-fit de Vaucouleurs profile
  with $r_e=100$~pc is shown as a dashed red line, and for comparison
  a broader profile with $r_e=0.04\arcsec=290$~pc is shown as the
  dotted green line.  This galaxy is quite compact, but is more
  extended than a point source.}
\label{fig:profile}
\end{center}
\end{figure}

For the most compact galaxy (J0905+5759, $r_e=0.013\arcsec$ or
100~pc), we also show the observed one-dimensional surface brightness
profile in Figure~\ref{fig:profile}.  We compare to the profiles of
six stars in the same image, the best-fit de Vaucouleurs model, and a
de Vaucouleurs model with $r_e=0.04\arcsec$ (the native WFC3/UVIS
pixel size, which corresponds to a physical scale of 290~pc).  This
comparison illustrates that this galaxy, while only marginally
resolved with an $r_e$ that is $\sim20\%$ of the image FWHM, is
clearly more extended than a point source.

\begin{figure}[!t]
\begin{center}
\includegraphics[angle=0,scale=0.41]{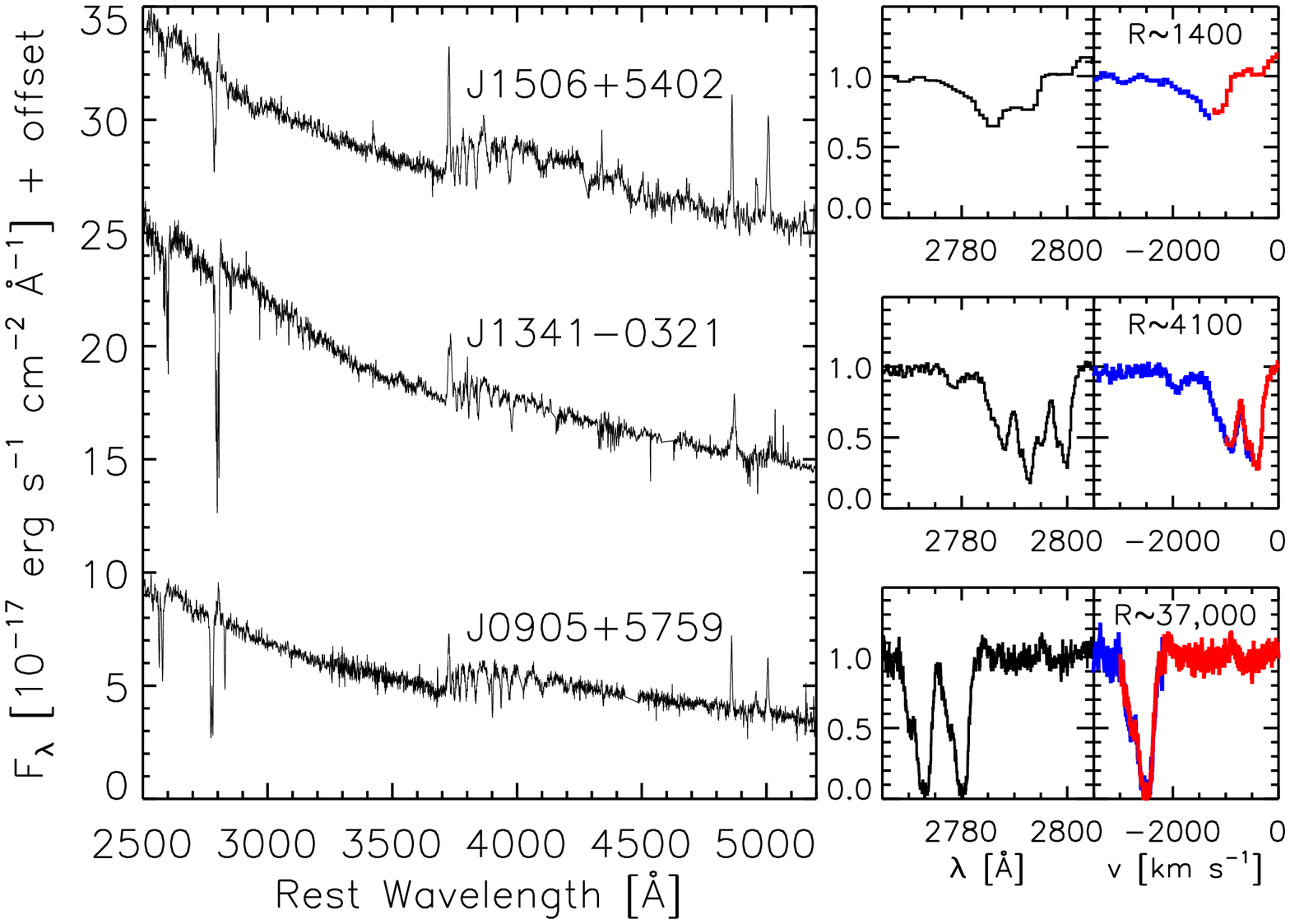}
\caption{Spectra covering $\lrest=2500$--5200~\AA\ for the three
  galaxies shown in Figure~\ref{fig:seds}.  For clarity the top two
  spectra are offset by $+10$ and $+20$ units.  These spectra are
  dominated by the light of a young stellar population but have
  relatively weak nebular emission lines (\oii~$\lambda3727$,
  H$\beta$~$\lambda4861$, \oiii~$\lambda5007$) and strong
  \mgii~$\lambda\lambda2796,2803$ absorption arising from the
  interstellar medium.  The absence of broad \mgii\ or H$\beta$
  emission lines rules out a significant contribution to the optical
  continuum light from an AGN.  The panels on the right highlight the
  region around the \mgii\ doublet in both wavelength and velocity
  space to illustrate the outflow kinematics.  The spectrum in the
  bottom panel has sufficient spectral resolution
  ($\textnormal{FWHM}\approx8$~\kms) to resolve the intrinsic shape of
  the absorption-line profile, revealing that the gas near the
  centroid velocity ($v=-2470$~\kms) covers the entire galaxy.}
\label{fig:spectra}
\end{center}
\end{figure}

For such a compact source, there is uncertainty in our $r_e$
measurement given uncertainties in the model PSF.  To quantify this,
we used TinyTim to generate a model PSF that is narrower than the
stars in the image (convolving with a $\textnormal{FWHM}=0.04\arcsec$
Gaussian, whereas the image FWHM is $0.074\arcsec$) and found that
this increased the $r_e$ in the GALFIT model by a factor of two.  We
also fit a two-component PSF+Sersic model, but found that the Sersic
component dominates the fit, yielding a similar $r_e$.  Furthermore,
the spectrum of the galaxy shows no evidence for an AGN contribution
to the optical continuum (see Figure~\ref{fig:spectra}), so there is
no clear motivation for including an unresolved, point-source
component in the model.  We conclude that our $r_e$ estimate is
accurate within a factor of two.

\subsection{Star Formation Rates and Stellar Masses}\label{sec:seds}

We gathered photometry from the WISE All-Sky Release, the SDSS Seventh
Data Release, and GALEX General Release 6.  We also obtained
$5\times30$~sec dithered exposures at 3.6~$\mu$m and 4.5~$\mu$m for
all sources with the Infrared Array Camera on the Spitzer Space
Telescope as part of GO program 60145.  We used the post--basic
calibrated data to perform aperture photometry on all sources and
point-source photometry on sources in crowded fields.  We show
spectral energy distributions (SEDs) for the three highest $\sigmasfr$
galaxies in Figure~\ref{fig:seds}.

We estimate IR-based SFRs for the 25/29 galaxies with WISE 12 or
22~$\mu$m detections by fitting \citet{cha01} templates to their 12
and 22~$\mu$m fluxes.  For the 14/25 galaxies with 22~$\mu$m
detections, this yields SFRs that agree with those obtained from the
\citet{ruj12} method based on 24~$\mu$m luminosity with a scatter of
0.05~dex.  Several authors have shown that the shape of the IR SED for
star-forming galaxies depends on $\sigmasfr$ \citep{ruj11,elb11}, with
more compact starbursts having larger total-IR (8--1000~$\mu$m) to
mid-IR (8--24~$\mu$m) ratios, characteristic of the most luminous
galaxies in the local universe \citep{rie09}.  If we used the most
luminous local templates for the 8/25 sources with SFRs in the ULIRG
regime ($\sfrir>100$~\msun~yr$^{-1}$), we would obtain SFRs that are
larger by 0.5~dex than the values we adopt for this Letter.

We also estimate SFRs and stellar masses based on stellar population
fits to the $\lrest=0.1$--3~$\mu$m SEDs using the method of
\citet{mou11}.  For the 14/25 galaxies with
$\sfrir>50$~\msun~yr$^{-1}$, there is agreement between these UV-based
SFR estimates and $\sfrir$ with a scatter of 0.32~dex.  For an SMC
dust law, we find a median attenuation of $A_V=0.4$~mag.  The observed
H$\beta$ luminosities, uncorrected for dust extinction, are typically
factors of 10--$20\times$ smaller than expected from the UV and IR
SFRs.  This can be reconciled by either strong differential dust
attenuation (i.e., $A_V\approx2$--3~mag for the line-emitting region),
escaping ionizing photons from matter-bounded \hii\ regions, or a
recently quenched starburst ($t>5~$Myr) with a small ratio of ionizing
($\lambda<912$~\AA) to non-ionizing UV photons.

\begin{figure*}[!t]
\begin{center}
\includegraphics[angle=0,scale=.82]{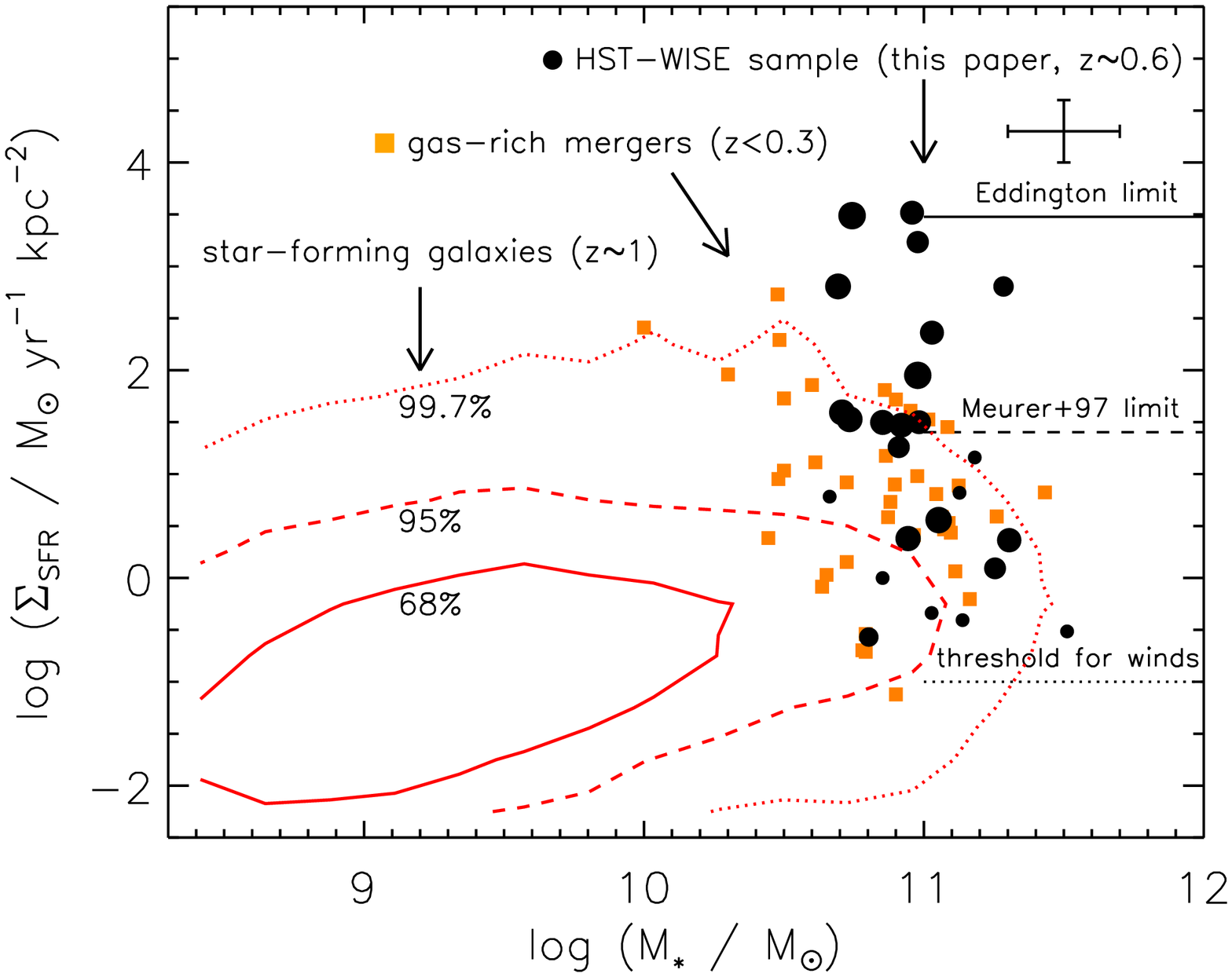} 
\caption{SFR surface densities and stellar masses for the HST--WISE
  sample described in this paper (black circles, symbol size
  proportional to outflow velocity), along with samples of $z<0.3$
  gas-rich mergers \citep[orange squares, including ULIRGs, Lyman
    break analogs, and Arp 220;][]{ken98,vei06,rod08,ove09} and
  $z\sim1$ star-forming galaxies \citep[shown with 68\%, 95\%, and
    99.7\% contours;][]{wuy11}.  We mark the empirical threshold for
  launching winds \citep[dotted line,
    $\sigmasfr\approx0.1$~\units;][]{hec02}, the 90th-percentile
  starburst intensity limit from \citet{meu97} (dashed line,
  $\sigmasfr\approx25$~\units), and the Eddington limit from radiation
  pressure on dust grains \citep[solid line,
    $\sigmasfr\approx3000$~\units;][]{mur05,tho05,hop10}.  The
  representative error bar in the top-right portion of the plot
  corresponds to uncertainties of 0.3~dex in $\sigmasfr$ and 0.2~dex
  in stellar mass.  Our HST--WISE sample overlaps with the region
  characterized by gas-rich mergers, and extends to very large SFR
  surface densities near the Eddington limit, suggesting growth that
  is limited by momentum injection from massive stars.}
\label{fig:sigmasfr}
\end{center}
\end{figure*}

\subsection{Outflow Kinematics and Covering Factors}

We present $\lrest=2500$--5200~\AA\ spectroscopy for three
high-$\sigmasfr$ sources in Figure~\ref{fig:spectra} based on data
from MMT/Blue Channel and SDSS (J1506+5402), Magellan/MagE
(J1341-0321), and Keck/LRIS and Keck/HIRES (J0905+5759).  The spectra
are dominated by light from a young ($t<50$~Myr) stellar population.
We highlight the interstellar medium \mgii~$\lambda\lambda2796,2803$
absorption lines, which are used to measure outflow velocities.  At
low spectral resolution (e.g., the top right panel of
Figure~\ref{fig:spectra}) it is not possible to determine the
intrinsic shape of the absorption line profile and therefore the
covering factor of the outflowing gas.  However, the Keck/HIRES
spectrum of J0905+5759 ($\textnormal{FWHM}\approx8$~\kms) reveals that
the gas covers the entire continuum source near the velocity centroid
($v=-2470$~\kms) indicating a galaxy-wide outflow.

\section{Discussion}\label{sec:discussion}

The compact sizes ($r_e\approx100$~pc) and large SFRs
($\textnormal{SFR}\approx200$~\msun) for the three galaxies described
above imply extremely large SFR surface densities
($\sigmasfr\approx3000$~\units).  To place these galaxies in context,
we plot $\sigmasfr$ versus stellar mass for the 25/29 galaxies
detected by WISE in Figure~\ref{fig:sigmasfr}.  We include comparison
samples of $\sim10^5$ star-forming galaxies at $0.5<z<1.5$ from
\citet{wuy11} and gas-rich mergers at $z<0.3$ including 32 ULIRGs from
\citet{vei06}, six Lyman break analogs with dominant central objects
from \citet{ove09}, and the local compact starburst Arp 220
\citep{ken98, rod08}.  We also mark the empirical threshold for
launching winds \citep[$\sigmasfr\approx0.1$~\units,][]{hec02}, the
90th percentile limit for the surface brightness of starbursts over a
wide range in redshift measured using UV, H$\alpha$, far-IR, and radio
continuum emission \citep[$\sigmasfr\approx25$~\units\ for a Chabrier
  IMF,][]{meu97}, and the theoretical limit for a starburst limited by
feedback from radiation pressure
\citep[$\sigmasfr\approx3000$~\units,][]{mur05,tho05,hop10}.  The most
luminous, compact starbursts in our sample exhibit SFR surface
densities that reach the Eddington limit, suggesting that their growth
is being regulated by momentum input from massive stars.

\subsection{Constraints on Ongoing AGN Activity}

While the SEDs (Figure~\ref{fig:seds}) and optical spectra
(Figure~\ref{fig:spectra}) for our sample indicate that their
$\lrest=0.1$--15~$\mu$m emission is dominated by star formation, it is
worthwhile to consider the level of ongoing AGN activity and its
effect on our measurements.  We observed 12/29 galaxies with the
Chandra X-ray Observatory (proposal ID 11700896, see P. Sell et al.,
in preparation for more details), including the 3/29 galaxies with
significant \nev~$\lambda3426$ detections \citep[an indication of AGN
  activity,][]{gil10} and the most luminous \oiii~$\lambda5007$
sources in the sample.  Of these 12 galaxies, only two were detected
with $>4$ X-ray counts, and upper limits for the remainder imply
$L_{\textnormal{\scriptsize{X}}}\lesssim10^{43}$~erg~s$^{-1}$.  The
two significant detections were for galaxies hosting optical
broad-line AGNs; neither was detected at 22~$\mu$m by WISE or shows
evidence for an outflow, and both have $\sigmasfr<10$~\units.  For the
most luminous \oiii\ source, J1506+5402, the observed $\oiii$/X-ray
ratio implies X-ray absorption by a factor of $\sim10$ \citep{hec05}.
However, even if the X-ray attenuation were a factor of 100, typical
for local Compton-thick AGNs \citep{dia09}, the expected mid-IR AGN
contribution \citep{gan09} would still be $\lesssim$30\% of the
observed mid-IR luminosity.  Only 1/29 galaxies has WISE mid-IR colors
that would identify it as an AGN candidate \citep{ste12}, and this
particular galaxy has $\sigmasfr<50$~\units.  We conclude that the
bolometric output of the galaxies in our sample is dominated by star
formation and that our results regarding large $\sigmasfr$ values are
not affected by AGN contamination.

\subsection{The Outflow Launching Mechanism}\label{sec:launch}

Are the high-velocity outflows observed in these galaxies produced by
a compact starburst?  \citet{mur11} argued that massive star clusters
with large gas surface densities are the ideal launching point for
galactic-scale outflows driven by radiation pressure, and that the
outflow velocity should scale with escape velocity of the most massive
star clusters in a galaxy.  For our sample, if one assumes that the
spatial extent of the stellar mass is comparable to that of the
rest-frame $V$-band light (see Section~\ref{sec:context}), then half
of the stellar mass ($\textnormal{\mstar}\sim10^{11}$~\msun) will be
within the effective radius ($r_{e}\sim100$~pc).  Such a compact
stellar population would have an escape velocity comparable to the
$\gtrsim1000$~\kms\ outflow velocities we observe:
\begin{eqnarray}\nonumber
v_{esc}&=&\sqrt{2GM_{*}/r} \\
&=& 2100
\left({M_{*}\over10^{11} M_{\odot}}\right)^{1/2} 
\left({r\over200~\textnormal{pc}}\right)^{-1/2}
\textnormal{km~s}^{-1} 
\label{eq:vesc}
\end{eqnarray}
This argument, combined with the fact that we observe galaxies with
significant dust-obscured star formation and $\sigmasfr$ values near
the Eddington limit, suggests that momentum input from massive stars
in the form of radiation pressure is a viable mechanism for launching
these outflows.

In addition to radiation pressure, we also expect significant momentum
flux from stellar winds and supernovae.  For example, a starburst with
$\textnormal{SFR}\approx200$~\msun~yr$^{-1}$ would have radiation
pressure $L_{bol}/c\approx3\times10^{35}$~dyne and ram pressure
$\dot{p}\approx5$--$10\times10^{35}$~dyne from stellar winds and
supernovae \citep{lei92,lei99,vei05}.  Furthermore, \citet{hec11}
noted that such a large momentum injection
($\dot{p}\approx10^{35}$~dyne) from a small initial radius
$r_{0}\approx100$~pc could accelerate a cloud with column density
$N_{H}\approx10^{21}$~cm$^{-2}$ to a terminal velocity
$v_{\infty}\approx1800$~\kms.  Thus, the energetics of compact
starbursts are sufficient to produce the high-velocity outflows we
observe, and it is plausible that both radiation pressure on dust
grains and supernova ram pressure contribute to driving the winds.

\subsection{Placing These Galaxies in Context}\label{sec:context}

It is clear from Figure~\ref{fig:sigmasfr} that the high-$\sigmasfr$
galaxies in our sample constitute a rare population, suggesting that
they represent an unusual or short-lived phase.  Mergers of gas-rich
galaxies are a viable mechanism for producing compact starbursts
\citep{mih96}, and such gas-rich major mergers are rare at $z\sim0.6$
due to the decline in both the gas fraction \citep{tac10} and the
fractional major merger rate \citep{lot11} of galaxies since $z\sim2$.
Furthermore, our high-$\sigmasfr$ galaxies are caught in a particular
time interval where there is both vigorous star formation and strong
feedback.  The length of this phase may be set by the gas consumption
timescale or the timescale for feedback to suppress subsequent star
formation.  Based on the Kennicutt--Schmidt relation \citep{ken98}, a
compact starburst with $r_e\approx100$~pc,
$\textnormal{SFR}\approx200$~\msun, and
$\sigmasfr\approx3000$~\units\ would have a gas surface density of
$\Sigma_{gas}\sim10^{11}$~\msun~kpc$^{-2}$ corresponding to
$M_{gas}\sim3\times10^{9}$~\msun\ inside 100~pc, which would be
consumed on a timescale $\sim20$~Myr.  This scenario could be tested
with CO observations of molecular gas masses and kinematics for these
extreme galaxies.

Considering our full HST--WISE sample in Figure~\ref{fig:sigmasfr}, we
find values of $\sigmasfr$ spanning four orders of magnitude.  This
could be explained as an evolutionary sequence where the
high-$\sigmasfr$ galaxies represent the peak of the starburst when the
high-velocity outflows are launched, while the lower $\sigmasfr$
galaxies represent a subsequent, post-starburst phase.  It is
interesting to note that all 12/25 galaxies above the
$\sigmasfr\approx25$~\units\ limit from \citet{meu97} exhibit outflows
(with median centroid velocity $v=-1500$~\kms), while all 7/25
galaxies without detected outflows (the smallest black circles in
Figure~\ref{fig:sigmasfr}) have $\sigmasfr<20$~\units.  If a compact
starburst is a requirement for the production of high-velocity
outflows, it may be that the sources without outflows have not gone
through such a phase (see A. Robaina, et al., in preparation for a
discussion of the relationship between galaxy morphology, outflow
velocity, and stellar population age in this sample).

Finally, we consider the implications of our results for models of
massive galaxy formation.  Simulations of major galaxy mergers with
large gas fractions ($f_{gas}\sim50$\%) can produce
$\textnormal{\mstar}\sim10^{11}$~\msun\ remnants with $r_{e}\sim1$~kpc
\citep{wuy10}, but our sample includes galaxies of similar mass that
are smaller in the rest-frame $V$ band by almost an order of
magnitude.  If the mass in these galaxies follows their $V$-band
light, it would be extremely challenging for them to grow in size from
$r_e\sim0.1$~kpc to $r_e\sim5$~kpc to reach the local size--mass
relation \citep{she03} in the $t\sim6$~Gyr since $z=0.6$.  For
comparison the compact, quiescent galaxies observed at $z\sim2$
\citep{tru07,van08} have $t\sim10$~Gyr to grow by a factor of $\sim5$.
In this context, it is worth noting that the half-light radius
(ignoring dust attenuation) at rest-frame $V$ band can be a factor of
5--10 smaller than the half-mass radius for a gas-rich merger at final
coalescence near the peak of starburst activity \citep{wuy10}.  One
could test for such size discrepancies and probe the radial dependence
of the mass-to-light ratio for these galaxies by measuring sizes at
rest-frame near-IR wavelengths.

\section{Summary}

We have measured large SFR surface densities for galaxies that exhibit
$\gtrsim1000$~\kms\ outflows.  The largest $\sigmasfr$ values are
comparable to the Eddington limit from radiation pressure on dust
grains, and such compact starbursts are expected to have substantial
momentum input from massive stars and supernovae.  High-velocity
outflows have been previously interpreted as a signpost of AGN
feedback, but given that feedback from a compact starburst is capable
of producing such a signature and is clearly present in this sample,
we conclude that the outflows we observe are likely driven by star
formation.

\acknowledgments

We acknowledge useful discussions with and assistance from James Aird,
Brandon Kelly, Dusan Keres, David Law, Alexander Mendez, Kate Rubin,
Art Wolfe, and Stijn Wuyts.  We thank the referee for suggestions that
have improved the paper.  AMD acknowledges support from the Southern
California Center for Galaxy Evolution, a multi-campus research
program funded by the University of California Office of Research.
Support for HST-GO-12272 was provided by NASA through a grant from
STScI.  Support for Spitzer-GO-60145 was provided by contract 1419615
from JPL/Caltech.  This Letter includes data obtained at the W.M. Keck
Observatory.

\end{document}